\documentstyle[colordvi,psfig,aps,preprint]{revtex}

\begin{document}
\title
{\bf Effect of baryon density on parton production, chemical equilibration and
thermal photon emission from quark gluon plasma}
\author{ D. Dutta, A. K. Mohanty, K. Kumar and R. K. Choudhury}
\address{ Nuclear Physics Division, Bhabha Atomic Research Centre, Mumbai, India-400085.}
\date{\today}
\maketitle
\begin{abstract}
The  effect of baryon density on parton production  processes of
$gg\rightleftharpoons ggg$ and $gg\rightleftharpoons q{\bar q}$
is studied using full phase space distribution function and also with inclusion
of quantum statistics i.e. Pauli blocking and Bose enhancement factors,
 in the case of both saturated and unsaturated quark gluon plasma.
The rate for the process $gg \rightleftharpoons q{\bar q}$ is found to be
much less as compared to the most commonly used factorized result obtained
on the basis of classical approximation.
This discrepancy, which is found both at zero as well as at finite baryon
densities, however, is not due to the lack of quantum statistics in the classical
approximation, rather due to the use of Fermi-Dirac
and Bose-Einstein distribution functions for partons instead of Boltzmann
distribution which is appropriate under such approximation.
Interestingly, the rates of parton production are found to be insensitive to the
baryo-chemical potential particularly when the plasma is unsaturated although
the process of chemical equilibration strongly depends on it.
The thermal photon yields,
have been calculated specifically from unsaturated
plasma at finite baryon density. The exact results obtained numerically are
found to be in close agreement with the analytic expression derived using
factorized distribution functions  appropriate for unsaturated
plasma. Further, it is shown that
in the case of unsaturated plasma, the thermal photon production is enhanced with
increasing baryon density both at fixed temperature and fixed energy density
of the quark gluon plasma.\\
PACS number(s):25.75.Dw, 12.38.Mh, 12.38.Bx, 24.85.+p
\end{abstract}
\vspace{.3cm}
{\bf ~~~~~~~~~~Submitted to Phys. Rev. C}\\
\section{ Introduction}
The future collider experiments with heavy ions will provide an
opportunity to investigate the physics of strongly interacting matter
at extreme energy densities where the formation of a new state of matter
, the quark gluon plasma (QGP), is expected. With these colliders,
it may be possible to achieve
energy density well above the deconfinement threshold  to be able to  probe the QGP in its
asymptotically free 'ideal gas' form. At high energies, the central
rapidity region may have nearly vanishing baryon density, similar to the early
universe, and will  probably consist of  a very dense system of semi-hard
partons (mini-jets) which would lead to rapid thermalization and extremely
high initial temperature \cite{1}. In the standard scenario, the quark
gluon plasma formed during the collision is expected to thermalise
in a typical time scale of
$(\approx 1$ fm/c) and the subsequent evolution  follows  the ideal Bjorken's
scaling solution \cite{2}.
Several theoretical calculations based on perturbative QCD approaches now
suggest that
the partonic fluid which is mostly gluonic will attain kinetic equilibrium
after a proper time $\tau_0 \approx 0.1-0.3 $ fm/c [3]. This is justified
given the fact that the rate of elastic collisions between quarks and gluons
which establish thermal and mechanical equilibrium (referred as kinetic
equilibrium corresponding to isotropic momentum distribution) is much larger
than for inelastic collisions which establish chemical equilibrium.
The remaining question is whether the high energy density matter remains in
QGP phase sufficiently long enough to actually reach chemical equilibrium so that
it can be identified with the QGP phase seen in lattice QCD calculations.

In the framework of pQCD based models, it is
now understood that the densities of quarks and anti-quarks may stay well below
the gluon density and may not reach up to the full equilibration value as
required by an ideal chemical mixture of quarks and gluons \cite{3}. Because of
the consumption of energy by additional parton production, the temperature
of the parton plasma falls down faster than the ideal solution where $T^3\tau$
remains constant. Essentially, same picture emerges in a somewhat different approach
first proposed by Biro et al \cite{4} and subsequently by several
other authors \cite{5,6,7}. With a given initial value for the energy and number
densities, this type of calculations employ ideal fluid dynamics to study the
subsequent evolution of the kinetically equilibrated QGP phase coupled to rate
equations to determine the chemical composition.

As mentioned before, the
evolution of the partonic matter toward a fully (chemically) equilibrated QGP
will be dictated by the parton proliferation mainly through induced radiation
and gluon fusion \cite{4,8}. Unfortunately, there have been no rigorous thermal
field theoretical calculations for the rate of quark production since the basic
cross section diverges for the mass less quarks. The closest approach to a full
calculation has been done by using  the thermal quark mass and Debye screening mass
to regularize the divergence \cite{4,9}. However, as used by Biro et al \cite{4}
and subsequently by many others \cite{5,6,7}, the quark production rate $gg \rightleftharpoons q{\bar q}$ has
been factorized using classical approximation where Pauli blocking and Bose
enhancement factors (effect of quantum statistics) have been eliminated.
The calculations have also been simplified further by assuming an ideal baryon
free plasma. There is now enough evidence from theoretical studies that even
at RHIC energies, the mid rapidity region may not be completely baryon free \cite{10}.

In an earlier work, we had studied chemical equilibration of the QGP at finite
baryon density \cite{7} where we had considered the processes $gg \rightleftharpoons ggg$
and $gg \rightleftharpoons q{\bar q}$ as the two dominant channels contributing to
parton production. It is natural to think that the
presence of baryon density $(\mu)$ in quark and anti-quark
distribution functions may directly affect the process $gg \rightleftharpoons q{\bar q}$,
whereas  the process $gg \rightleftharpoons ggg$ may depend on $\mu$ only through
the Debye screening  mass used to regularize the divergence.
Using  modified Fermi-Dirac distribution function for quarks (anti-quarks)
and a factorized distribution function for gluons
valid at small baryon density, it was found  in the above work that the two rates are not
sensitive to the baryo-chemical potential particularly when the plasma is unsaturated.
Therefore, we had used the same factorization (for $gg \rightleftharpoons q{\bar q}$)  as
used by Biro et al in the case of baryon free plasma. Although, the parton production rates
for unsaturated plasma are not very sensitive to baryo-chemical potential, the energy,
quark and anti-quark densities (hence the rate of chemical equilibration)
strongly depend on $\mu$. It was also shown that due to chemical equilibration, the
baryon rich plasma cools at a  slower rate as compared to the
baryon free plasma.

In the present  work, we have studied the role of baryon density on parton
production particularly for the process $gg \rightleftharpoons q{\bar q}$ in detail
using full phase space distribution functions for partons and also including Pauli blocking
and Bose enhancement factors explicitly. Specifically, we investigate
the effect of quantum statistics particularly in the presence of finite
baryon density. The present calculations
show that the rate for the process $gg \rightleftharpoons q{\bar q}$ is lowered
 by 20$\%$ to 30$\%$
as compared to the factorized result of Biro et al as given in Ref. \cite{4}.
This discrepancy, however, is not due to the
lack of quantum statistics, but rather due to the use of
Fermi-Dirac (FD) and Bose-Einstein (BE) type
distribution functions for quarks and gluons in Ref. \cite{4} instead
of Boltzmann distribution
which would have been appropriate under classical approximation.
With the revised rates, we have also studied the process of chemical
equilibration solving a set of rate equations.
The quark and anti-quark fugacities slow down
whereas the gluon fugacity goes up slightly as compared to the previous values.
However, there is no net effect on the evolution of
 temperature which remains practically unchanged. Further, it is
shown that the rates of parton production are insensitive to the presence of
baryon density particularly when the plasma is unsaturated. The findings of the
present study are also consistent with our earlier work \cite{7}. We
had earlier studied the  thermal dilepton yields
 as a  probe of QGP. We investigate here the thermal
hard photon production as a complementary probe from a chemically non-equilibrated baryon rich quark gluon plasma.
It is seen that, like the dilepton yields, the
photon yield is also enhanced in presence of baryon density.

The paper is organized as follows. In section II, we calculate the parton
production and chemical equilibration at finite baryon density.
 In section III, we  calculate the thermal photon yield from
 chemically unsaturated plasma both numerically and analytically. The
  results have been summarized in section IV.

\section{Parton production and chemical equilibration}
In this section, we calculate the thermal production rates for
 (i) gluon fusion $gg\rightleftharpoons   q{\bar   q}$ and (ii) gluon radiation
 $gg\rightleftharpoons  ggg$. We
 also study the process of chemical equilibration by solving the ideal
 fluid hydrodynamics  coupled to a set of rate equations.
 We use the Juttner distribution functions for quarks (anti-quarks) and gluons
 as given by
\equation f_ q(_{\bar q})  = \frac{\lambda_ q(_{\bar q})}
{\lambda_ q(_{\bar q}) + e^{(p \mp\mu)/T}}=
\frac{\lambda_{q({\bar q})}  e^{\pm x}}
{\lambda_{q({\bar q})} e^{\pm x}+ e^{p /T}} ~~~~
;~~~~~f_ g = \frac{\lambda_ g}{e^{p/T}-\lambda_g}
\endequation
where $x=\mu/T$. The fugacity factor $\lambda_i $ ($i=q,\bar{q}$ and g) gives the measure of
the deviation of the distribution functions from the equilibrium values
 and $\mu$ (= $\mu_B$/3) is the quark-chemical potential.
The chemical equilibrium is said to be achieved when $\lambda_i \rightarrow 1$.
As mentioned in Ref. \cite{7},  the quark (anti-quark)
distribution functions could also be written using the commonly used definition of fugacities
$\lambda_Q$ and $\lambda_{\bar Q}$
given by
\equation \lambda_{Q({\bar Q})}=e^{\pm x}\lambda_{q({\bar q})} \endequation
so that, there is no need to use the quark chemical potential $\mu$ explicitly.
However, the definition (1) is quite convenient, since  at
equilibrium where $\lambda_q$=$\lambda_{\bar q}$=1, the chemical potential associated with $\lambda_i$ vanishes
but the baryo-chemical potential still exists. Further,
we have taken $\lambda_{q} = \lambda_{\bar q}$ at all values of $\tau$ so that
when $\mu\rightarrow  0$, $\lambda_{Q} = \lambda_{\bar Q} =\lambda_q$ resulting in a baryon
symmetric matter.
While we calculate the parton production rates using the above distribution
functions, we also consider the approximations for quarks (anti-quarks) as
\equation
f_ {q({\bar q})} =
\frac{\lambda_{q({\bar q})}  e^{\pm x}}{\lambda_{q({\bar q})} e^{\pm x}+ e^{p /T}}
\approx \frac{\lambda_ {q({\bar q})} e^{\pm x}}{1+ e^{p /T}}= \lambda_{q({\bar q})} e^{\pm x} f^{eq}_{q (\bar{q})}
\endequation
and for gluon distribution function as
\equation  f_g=\lambda_g f_g^{eq} \endequation
where $f_{q({\bar q})}^{eq} = (1+e^{p/T})^{-1} $ and $f_g^{eq} = (e^{p/T}-1)^{-1} $.
The approximation given by Eq.(3) is referred as modified Fermi-Dirac (MFD)
distribution in Ref. \cite{7} which becomes Fermi-Dirac distribution when
$\lambda_{Q({\bar Q})} \rightarrow 1$.
Hence forth, we will refer both Eqs.(3) and (4) as modified factorized distributions
(MFD) which are different from Boltzmann (BM) type of factorized form given by,
\equation f_ q(_{\bar q})  = \lambda_ q(_{\bar q})e^{-(p \mp\mu)/T}
~;~~~f_g=\lambda_g~e^{-p/T}\endequation

\subsection{ Gluon fusion }
 We begin by calculating the rate for the process
$gg\rightleftharpoons q{\bar q}$ \cite{8}:

$$ R_{g\rightarrow q}= \frac{1}{2} \int \frac{d^3p_1}{(2\pi)^3 2 E_1}
\int \frac{d^3p_2}{(2\pi)^3 2 E_2} \int \frac{d^3p_3}{(2\pi)^3 2 E_3}
\int \frac{d^3p_4}{(2\pi)^3 2 E_4} (2\pi)^4 \delta^4(p_1+p_2-p_3-p_4) ~\times
$$
\equation
\sum |{\cal M}_{gg\rightarrow q\bar q}|^2 f_g(p_1)f_g(p_2)[1-f_q(p_3)][1-f_{\bar q}(p_4)]
\endequation
and
$$ R_{q \rightarrow g}= \frac{1}{2} \int \frac{d^3p_1}{(2\pi)^3 2 E_1}
\int \frac{d^3p_2}{(2\pi)^3 2 E_2} \int \frac{d^3p_3}{(2\pi)^3 2 E_3}
\int \frac{d^3p_4}{(2\pi)^3 2 E_4} (2\pi)^4 \delta^4(p_1+p_2-p_3-p_4)~\times
$$
\equation
\sum |{\cal M}_{q \bar q\rightarrow gg}|^2 f_q(p_3)f_{\bar q}(p_4)[1+f_g(p_1)][1+f_g(p_2)]
\endequation
In Eq.(6), the squared matrix element, summed over spin and color,
$\sum |{\cal M}|^2$ is weighted by two
gluon distribution functions $f_g$ for the initial states. The factor $
[1-f_q][1-f_{\bar q}]$ indicates Pauli blocking for the final states.
In the reverse process {Eq.(7)),
the rate is weighted by the distribution functions of quarks
and anti-quarks
for the initial states and the gluon final states each gain an enhancement
factor $[1+f_g]$ due to Bose-Einstein statistics. The factor of 1/2 accounts
for the identity of the two gluons. Using the identity that results from Eq.(1),
 \begin{equation} [1-f_{q({\bar q})}]= \frac {f_{q({\bar q})}}
 {\lambda_{Q({\bar Q})}} e^{p/T}~; ~~
 [1+f_g(p)]= \frac {f_g} {\lambda_g} e^{p/T} \end{equation}
and the unitary relation $|{\cal M}_{12}|^2=|{\cal M}_{21}|^2$, Eq.(6) and Eq.(7)
can be combined to give
\begin{equation}
R_{g\rightarrow q}- R_{q\rightarrow g}
= \left[\frac{1}{\lambda_Q\lambda_{\bar Q}}-\frac{1}{\lambda_g^2}\right]~ I_{gluon}
\end{equation}
where
\begin{eqnarray}
I_{gluon}&=& \frac{1}{2} \int \frac{d^3p_1}{(2\pi)^3 2 E_1}
\int \frac{d^3p_2}{(2\pi)^3 2 E_2} \int \frac{d^3p_3}{(2\pi)^3 2 E_3}
\int \frac{d^3p_4}{(2\pi)^3 2 E_4} (2\pi)^4 \delta^4(p_1+p_2-p_3-p_4)~\times \nonumber \\
&&\sum |{\cal M}_{gg\rightarrow q\bar q}|^2 f_g(p_1)f_g(p_2)~f_q(p_3)~f_{\bar q}(p_4)
\times e^{\beta~(E_1+E_2)}
\end{eqnarray} and $\beta^{-1}=T$.
The above multi-dimensional integral can be reduced
further  by replacing
\equation \int \frac{d^3p}{2E}=\int d^4p \delta(p^2-m^2)\theta(p_0)\endequation
and by change of variables to:
\begin{eqnarray}
q&=&p_1 + p_2, \nonumber\\
p&=&\frac{1}{2}( p_1 - p_2), \nonumber\\
q^\prime&=&p_3 + p_4, \nonumber\\
p^\prime&=&\frac{1}{2}( p_3 - p_4).
\end{eqnarray}
One can easily eliminate the energy-momentum-conserving $\delta$ function by carrying out the
integrals over $q^\prime$. Using spherical coordinates and defining $q$ along the
z-axis where
$q_\mu=(q_0,0,0,q),~~p_\mu = (p_0,p\sin\theta,0,p\cos\theta)$ and
$p^\prime_\mu = (p^\prime_0, p^\prime\sin\phi \sin\chi,
p^\prime\sin\phi \cos\chi,p^\prime\cos\phi)$,
we get,
\begin{eqnarray}
I_{gluon}&=&\frac{1}{16 (2\pi)^6}\int_{0}^{\infty}dq_0~\int_0^\infty
 dq ~\int_{-q_0/2}^{q_0/2}dp_0~\int_{-q_0/2}^{q_0/2} dp_0^\prime
 \int_{0}^{\infty} dp  \int_{0}^{\infty} dp^\prime
 \int_{-1}^{1} d(\cos \theta) \int_{-1}^{1} d(\cos \phi) \nonumber\\
 &&\times \int_0^{2\pi} d \chi\delta\left[ p- \left[p_0^2 + \frac{s}{4}\right]^{\frac{1}{2}}\right]
 \delta \left[ p^\prime -\left[{p^\prime_0}^2 - m^2  + \frac{s}{4}\right]^{\frac{1}{2}}\right]
 \delta\left[\cos\theta - \frac{q_0~p_0}{qp}\right]
 \delta\left[\cos\phi - \frac{q_0~p^\prime_0}{qp^\prime}\right]\nonumber\\
&&\times  \sum |{\cal M}_{gg\rightarrow q\bar q}|^2
f_g\left[\frac{q_0}{2}+p_0\right]~f_g\left[\frac{q_0}{2}-p_0\right]
~f_q\left[\frac{q_0}{2}+p_0^\prime\right]
f_{\bar q}\left[\frac{q_0}{2}-p_0^\prime\right]~e^{\beta q_0}.
\end{eqnarray}
The summed squared matrix element for the process $gg\rightarrow q{\bar q}$,
is given by \cite{8} (where
quarks are assumed to  be massless)
\equation
\sum |{\cal M}_{gg\rightarrow q{\bar q}}|^2=\gamma_g^2~\gamma_q\gamma_{\bar q}~\pi^2~\alpha_s^2
\left[\frac{ut}{3s^2}+\frac{2}{27}(\frac{u}{t}+\frac{t}{u})-\frac{1}{6}\right].
\endequation
Here $\gamma_g\equiv 2(N_c^2-1)$ and $\gamma_q=\gamma_{\bar q}\equiv 2N_c N_f$
are the number of internal degrees of freedom for gluons, quarks and anti-quarks
respectively.  $N_c=3$ is the number of colors and $N_f$ is the number
of massless flavors. Throughout our analysis we have used $N_f=2.0$.
The Mandelstam variables $s,~t,~u$ used in the above equations are given by
\begin{eqnarray}
s&=&q_0^2-q^2,\\
t&=&m^2-\frac{s}{2}\left(1-\frac{4p_0~p_0^\prime}{q^2}\right)+2
\left[\frac{s}{4}\left(1-\frac{4p_0^2} {q^2}\right)\right] ^{\frac{1}{2}}
\left[\frac{s}{4}\left(1-\frac{4{p_0^\prime} ^2}{q^2}\right)-m^2\right]
^{\frac{1}{2}}\sin\chi,\\
u&=&-(s+t).
\end{eqnarray}
The matrix element  given by Eq. (14)  diverges in the limit $u,t \rightarrow 0$.
Hence, the  medium induced effective quark mass plays a crucial role and
the divergence of the cross-section can be avoided by replacing 'm'
with the thermal quark mass defined appropriately for
a non equilibrium plasma \cite{11,12}
\equation
m=m_q^2 = \frac{~g^2}{3\pi^2} \int_0^\infty dp~p~[2~f_g+f_q+f_{\bar q}].
\label{qmass}
\endequation

The integrals over  $p, p^\prime, \theta$ and $\phi$ of Eq. (13) have been performed
carefully because of kinematical constraints to have the limits
\equation
q_0\geq 2m,~~q^2\leq (q_0^2-4m^2), ~~p_0^2\leq \frac{q^2}{4},~~
{p_0^\prime}^2\leq \frac{q^2}{4}\left[1-\frac{4m^2}{s}\right].
\endequation
Thus we evaluate all but five of the integrals in Eq.(13) trivially \cite{8}
obtaining
\begin{eqnarray}
I_{gluon}&=& \frac{1}{16 (2\pi)^6}\int_{2 m}^{\infty}dq_0~\int_0^{(q_0^2-4m^2)
^{\frac{1}{2}}} dq ~\int_{-p_*}^{p_*}dp_0~\int_{-p_*^\prime}^{p_*^\prime} dp_0^\prime \nonumber\\
&&\times \int_0^{2\pi} d \chi \sum |{\cal M}_{gg\rightarrow q\bar q}|^2
f_g[\frac{q_0}{2}+p_0]~f_g[\frac{q_0}{2}-p_0]~f_q[\frac{q_0}{2}+p_0^\prime]
f_{\bar q}[\frac{q_0}{2}-p_0^\prime]~e^{\beta q_0}
\end{eqnarray}
where the limits of integration are
\begin{eqnarray}
p_*&=&\frac{q}{2},  \\
p_*^\prime&=&\frac{q}{2}(1-\frac{4m^2}{s})^{\frac{1}{2}}.
\end{eqnarray}
Following Ref. \cite{4}, we write Eq.(9) in a convenient form
\begin{equation}
R_{g\rightarrow q}- R_{q\rightarrow g}
= n_gR_2(1-\frac{\lambda_Q\lambda_{\bar Q}}{\lambda_g^2})
\end{equation}
where $n_g$ is the gluon density and $R_2$ is given by
\begin{equation}
R_2= \frac{I_{gluon}}{\lambda_Q\lambda_{\bar Q}n_g}.
\end{equation}
\begin{figure}[h!]
\vspace{6cm}
\hspace{1cm}
\begin{minipage}[t]{10cm}
\psfig{figure=fig1.out,width=10cm,height=10cm}
\end{minipage}
\caption{(a)The quark production rate $R_2/T$ as a function of
      $\lambda_g$ ($\lambda_q=\lambda_{\bar q}=\lambda_g$)
      with Juttner, Boltzmann and MFD distribution function.
      for $x=0$. The  solid dot curve is the approximate
      formula used by Biro et al.[4].The dashed dot curve
      is calculated rate with Boltzmann distribution function
      in the incoming channel and factorised distribution
      otherwise.
      (b)Same as (a) but with $(\lambda_q =\lambda_{\bar q}= \lambda_g/5)$.
      (c) The quark production rate $R_2/T$ as a function of $x$ for
      different values of $\lambda_g$ where $(\lambda_q =\lambda_{\bar q}= \lambda_g/5)$.}
\label{Figure 1}
\end{figure}
We compute $R_2$ numerically using Juttner, MFD and BM distribution functions
and also compare the result  with the approximation obtained by Biro et al
\begin{equation}
R_2 \approx 0.24 N_f~\alpha_s^2~ \lambda_g ~T ~ln(\frac{1.65}{\alpha_s \lambda})
\end{equation}
where $\alpha_s(=0.3)$  is the strong coupling constant.
Note that the factor $\lambda= (\lambda_g+\cosh x\lambda_q/2)$ which
arises due to the thermal quark mass \cite{7} is slightly different
from what is used in Ref. \cite{4}

Figs. 1(a) and 1(b) show $R_2/T$ as a function of $\lambda_g$ obtained numerically
using Eq.(24) with Juttner, MFD and BM distribution functions at T =0.57 GeV. Initially, we consider
a baryon free plasma ($x=0$). In Fig. 1(a), we assume the gluon and quark (anti-quark)
content of the plasma is equal ($\lambda_q=\lambda_g$), whereas in Fig. 1(b),
the plasma is assumed to be more gluon rich (say $\lambda_q=\lambda_g/5$). As
in Fig. 1(a), the $R_2/T$ values obtained using Juttner, MFD and BM distribution functions
do not have significant difference at all values of $\lambda_g$. At small $\lambda_g$,
$R_2/T$ obtained using MFD differs slightly which ultimately merges with that obtained using
Juttner distribution function, as both Juttner and MFD distributions become equal when $\lambda_q=\lambda_g=1.0$
at $x=0$. However, when the plasma is gluon rich as in Fig. 1(b), $R_2/T$ values
obtained with above three distributions do not differ much say up to $\lambda_g\approx 0.5$,
beyond which the MFD and BM results show significant deviation. Recall that for
$\lambda_q=\lambda_g/5$, MFD and Juttner distributions (for $q$ and $\bar q$) are not equal when
$\lambda_g \rightarrow 1$ even at $x=0$. We can conclude from these results that
for unsaturated plasma, the $R_2$ values obtained using
Juttner distribution do not deviate significantly from the results that can be obtained
with the factorized distributions like MFD or BM functions. However, these
results are quite less (20$\%$ to 30 $\%$) than what is found on the basis of
the so called classical approximation i.e. using Eq.(25) given by  Biro et al
which are also shown in Figs. 1(a) and 1(b) for comparison.  It is
natural to think that the discrepancy found in Biro et al's estimation
could be  due to the classical aspect of the approximation which we investigate
below.

In the classical approximation, Eq.(10) can be written as
\begin{equation}
I_{gluon} = \frac{1}{2} \int \frac{d^3p_1}{(2\pi)^3 }
\int \frac{d^3p_2}{(2\pi)^3 }
[\sigma_{gg \rightarrow q\bar q} v_{12}] f_g(p_1) f_g(p_2)
\end{equation}
which represents the cross section for the process $ gg \rightarrow q\bar q$
folded with the distributions for the initial particles. The cross section
$\sigma_{gg \rightarrow q\bar q}$ is given by
\begin{eqnarray}
\sigma_{gg \rightarrow q\bar q}&= &\frac{1}{v_{12} 2E_1 2E_2}
\int \frac{d^3p_3}{(2\pi)^3 2 E_3} \int \frac{d^3p_4}{(2\pi)^3 2 E_4}
(2\pi)^4 \delta^4(p_1+p_2-p_3-p_4)
 \sum |{\cal M}_{gg\rightarrow q\bar q}|^2
\end{eqnarray}
The classical approximation \cite{7} assumes Boltzmann distribution function
for quark, anti-quark and gluon and also eliminates the Pauli blocking and
Bose enhancement factors in the final states. The Eq.(26)
can be factorized as $I_{gluon}={\bar n_g^2}\sigma_2/2$ where ${\bar n_g}$ is the
equilibrium gluon density and $\sigma_2$ is a velocity weighted cross section.
Defining $R_2=n_g\sigma_2/2$, Biro et al estimate $\sigma_2$, $n_g$ and also thermal quark mass
using factorized distribution functions ( same as Eqs. (3) and (4)  with $x$=0 which
become Fermi-Dirac and Bose-Einstein distributions in the limit $\lambda_i \rightarrow 1.0$).
 Within the
present formalism, we can simulate the results of Biro et al by replacing $f_q$
and $f_{\bar q}$ in Eq.(20) with BM distribution (Eq. 5) and $f_g$ with
Bose-Einstein type (Eq. 4). Also we estimate thermal quark mass using MFD type
distributions as in Ref. \cite{4} (i.e. using Eqs. (3) and (4) at $x$=0). The simulated
result is also shown in Figs. 1(a) and 1(b) for comparison (see solid curve with dots).
The results are in good agreement with that of Biro et al's estimation. Therefore,
the deviation found using Eq.(25) is due to the use of FD and BE type of distribution
functions instead of using BM functions which would have been appropriate under
classical approximation. In fact, our results  obtained from Eq.(20)
using BM approximation is in a way classical.
Further it may be mentioned here that the first term in the RHS of Eq.(9) arises
due to the factors like Pauli blocking $(1-f_q)$ and Bose enhancement $(1+f_g)$.
Only the integral $I_{gluon}$ is evaluated using various approximations. Therefore,
even though we use the classical approximation like BM distribution, the final
results still include the quantum effect. Similarly, Biro et al have estimated $I_{gluon}$
classically, but their final factorization does include quantum effect (see appendix B
of Ref. \cite{7} for detail). In other words, Biro et al approach will give
correct estimation if BM approximation is used to estimate $R_2$.
The above discussions confine only to baryon free plasma.
Similar discrepancy is found at finite baryon density also.

Now we investigate the
effect of baryon density on $R_2$. Fig. 1(c) shows $R_2/T$ as a function of $x$
obtained numerically using actual (Juttner) distribution function at different values of
fugacities where $\lambda_q=\lambda_g/5$. It is interesting to note that $x$
dependence of $R_2$ is rather weak particularly when the plasma is unsaturated.
The nearly $x$ independence is also evident from
Eq. (1).
For small values of $\lambda_q$, the contribution from the factor $\lambda_q e^{\pm x}$
in the denominator is not very significant if $x$ is small. The $x$
dependence of the quark and anti-quark distribution functions mainly arises
due to the $e^{\pm x}$ factor in the numerator. Since these exponential
factors get cancelled in the product, $f_q f_{\bar q}$ (hence  $R_2$) will have weak
$x$ dependence at small baryon density if the plasma is highly unsaturated.

\subsection{ Gluon multiplication }

 The rate $R_3$ for the process $gg \rightleftharpoons ggg$ depends on the differential
radiative cross section \cite{4,5}
\equation
\frac{d\sigma_3}{dq_\perp^2~dy~d^2k_\perp} =
\frac{d\sigma_{el}^{gg}}{dq_\perp^2}\frac{dn_g}{dy~d^2k_\perp}
\theta(\lambda_f- \frac{\cosh y}{k_\perp})~\theta({\sqrt s} - k_\perp~\cosh y) \endequation
where the first step function gives approximate LPM suppression factor
and the
second step function accounts for energy conservation, $s = 18~T^2$
is the average squared center-of-mass energy of two gluons in the thermal gas.
Here $k_\perp$ denotes the transverse momentum, $y$ is the longitudinal rapidity of
the radiated gluon and $q_\perp$ denotes the momentum transfer in the elastic
collision.
The regularized gluon density distribution induced by a single scattering is
\equation
\frac{dn_g}{dy~d^2k_\perp}=\frac{C_A\alpha_s}{\pi^2}\frac{q_\perp^2}{k_\perp^2
[({\bf k_\perp-q_\perp})^2+m_D^2]}.
\endequation
Similarly the regularized small angle  $gg$ scattering cross section is
\equation
\frac{d\sigma_{el}^{gg}}{dq_\perp^2}=\frac{9}{4}\frac{2\pi\alpha_s^2}{(q_\perp^2
+m_D^2)^2}\endequation
where Debye screening mass  $m_D$ given by
\equation
m_D^2 = \frac{3~g^2}{\pi^2} \int_0^\infty dp~p~[2~f_g+N_f(f_q+f_{\bar q})]
\endequation
where $g^2 = 4\pi\alpha_s$.
The mean free path $\lambda_f$ for elastic scattering is then \cite{5,7}
\equation
\lambda_f^{-1} = \frac{9}{2}~\pi~\alpha_s^2~n_g~
\frac{1}{m_D^2}\left[1+\frac{1}{(1+\frac{2}{9}\frac{m_D^2}{T^2})}\right]
\endequation

Integrating  $\phi$ part   analytically
\equation
\int_0^{2\pi}~d\phi \frac{1}{({\bf k_\perp-q_\perp})^2+m_D^2}=\frac{2\pi}{\sqrt
{(k_\perp^2+q_\perp^2+m_D^2)^2-4q_\perp^2k_\perp^2}},
\endequation
and defining $R_3=\frac{1}{2}n_g\sigma_3$,
we can evaluate
\equation
R_3/T = \frac{27\alpha_s^3}{2} \lambda_f^2~n_g~I(\lambda_g,\lambda_q,x)
\endequation
where $I(\lambda_g,\lambda_q,x)$ is a function of $\lambda_g,\lambda_q$ and $x$,
\begin{eqnarray}
I(\lambda_g,\lambda_q,x)& = &\int_1^{\sqrt s \lambda_f} dx \int_0^{s/4 m_D^2}
dz \frac{z}{(1 + z)^2}\nonumber\\
&&\left[ \frac{\cosh^{-1}(\sqrt {x})}
{ x \sqrt{[x + (1+z)x_D]^2-4 x z x_D}}
 + \frac{1}{s \lambda_f^2} \frac{\cosh^{-1}(\sqrt x)}
 {\sqrt {[1+x (1+z)y_D]^2-4 x z y_D}} \right]
\end{eqnarray}
with $x_D=m_D^2~\lambda_f$ and $y_D=m_D^2/s$.

As mentioned before, $R_3$ depends on baryo-chemical potential $\mu$ through
the Debye screening mass  $m_D$
which we evaluate using Juttner, BM and MFD distributions.
\begin{figure}[h!]
\hspace{3cm}
\begin{minipage}[t]{12cm}
\psfig{figure=fig2.out,width=12cm,height=12cm}
\end{minipage}
\caption{ (a)The gluon production rate, $R_3/T$
      as function of $\lambda_g$ $(\lambda_q =\lambda_{\bar q}= \lambda_g/5)$
      with Juttner, Boltzmann and MFD distribution function.
      for $x=0$.
      (b) The gluon production rate $R_3/T$ as a function of $x$ for
      different values of $\lambda_g$ where $(\lambda_q =\lambda_{\bar q}= \lambda_g/5)$. }
\label{Figure 2}
\end{figure}
Fig. 2(a) shows $R_3/T$ as a function of $\lambda_g$ for a typical
gluon rich plasma.
At small values of $\lambda_g$, $R_3/T$ is not much
sensitive on various distribution functions, but they start deviating when
$\lambda_g$ exceeds $\approx 0.5$. Since the plasma remains highly unsaturated
by the time $T$ drops to $T_c$, results based on MFD or BM distribution will
not differ much from the actual result. Fig. 3(b) shows the variation of $R_3/T$
with $x$ at few typical values of $\lambda_g$. Again, the $x$ dependence on
$R_3$ is found to be insignificant particularly at small baryon density.

\subsection{ Chemical equilibration}

The  space time  evolution  of  the  QGP is studied using ideal hydrodynamics
in (1+1) dimension along with the following master rate equations \cite{4}
\equation \frac{\partial n_g}{\partial \tau} = (R_{2\rightarrow 3} -R_{3\rightarrow 2})
-(R_{g\rightarrow q} - R_{q\rightarrow g})\label{rate1} \endequation
\equation \frac{\partial n_q}{\partial \tau} = \frac{\partial n_{\bar q}}{\partial \tau}=
(R_{g\rightarrow q} - R_{q\rightarrow g})\label{rate2}\endequation
where $R_{2\rightarrow 3}$ and $R_{3\rightarrow 2}$ denote the rates for the
process
$gg \rightarrow ggg$ and its reverse and $R_{g \rightarrow q}$ and $R_{q \rightarrow g}$
for the process $gg \rightarrow q \bar q$ and its reverse respectively.
Similarly, the equation for the conservation of energy and momentum in (1+1) dimension can be
written as
\begin{equation}   \frac {\partial \epsilon }{ \partial \tau} + \frac {\epsilon+p} {\tau }=0 \end{equation}
which does not include  viscosity effect \cite{13}.
In case of an ideal fluid, 3p =$\epsilon$. From the conservation of baryon
number, one gets $\partial_\mu(n_Bu^\mu)=0$ which results in
\equation
n_B\tau=(n_q-n_{\bar q})\tau=const\endequation

We evaluate the RHS of Eq.(36) and Eq.(37) numerically using Juttner distribution.
However, in order to solve the above coupled equations, we use  factorized
density distributions $n_g=\lambda_g {\bar n_g}$ and $n_{q(\bar q)}=
\lambda_{Q(\bar Q)}{\bar n_{q({\bar q})}}$ (where $\bar n_i$ are the equilibrium parton densities)
obtained on the basis of Eqs. (3) and (4). Since the plasma
is unsaturated, we could have also used the BM approximation (Eq. 5). But, we prefer
to use Eqs. (3) and (4) for consistency with earlier works \cite{4,7}.

Finally, we solve the following set of equations for
 $\lambda_g$, $\lambda_Q$,
$\lambda_{\bar Q}$ and $T$ numerically by using fourth order Runga-Kutta method \cite{7},
\begin{equation} \frac{{\dot \lambda}_g}{ \lambda_g }+ \frac{ {3 \dot T}}{T} + \frac{1}{\tau} =
         R_3(1-\lambda_g)-2 R_2 ( 1- \frac{\lambda_Q  \lambda_{\bar Q} }
         { \lambda_g^2}), \end{equation}

\begin{equation} \frac{{\dot \lambda}_Q}{ \lambda_Q }+ \frac{ {3 \dot T}}{T} + \frac{1}{\tau} =
         R_2 \frac{ a_1}{b_1} \frac{\lambda_g}{ \lambda_Q} ( 1- \frac{\lambda_Q  \lambda_{\bar Q} }
         { \lambda_g^2}), \end{equation}

\begin{equation} {\dot \lambda}_{\bar Q} ={\dot { \lambda}_Q }+
                   \frac { \lambda_Q -  \lambda_{\bar Q }}{\tau}    +
         \frac{ 3 {\dot T}}{T} (\lambda_Q -  \lambda_{ \bar Q}),
 \end{equation}

 \begin{equation} \frac {3 {\dot T}} {T}+ \frac{1}{\tau} = -~~\frac{3}{4~A_t}
                  [  a_2 {\dot \lambda}_g +  b_2 {\dot \lambda}_Q
                  +  b_2 {\dot \lambda}_{\bar Q}   ],
 \end{equation}

where $$ A_t =  a_2 \lambda_g + b_2 (\lambda_Q + \lambda_{\bar Q}). $$

As in Ref. \cite{5,7}, we take the initial conditions as
$T_0$= 0.57 GeV, $\lambda_{g0}$=0.09, $\lambda_{q0}$=0.02 at
$\tau_0$ =0.3 fm/c  and treat $x_0$($\mu/T$) as a parameter.
\begin{figure}[h!]
\hspace{3cm}
\begin{minipage}[t]{12.0cm}
\psfig{figure=fig3.out,width=12.0cm,height=12.0cm}
\end{minipage}
\caption { (a) The temperature,  the quark chemical potential $x=\mu /T$,
       the gluon fugacity $\lambda_g$, the quark fugacity $\lambda_q$,
      as a function of $\tau$ for $R_2$ integrated using Juttner
      distribution and also for $R_2$ using approximate formula
      of Biro et al. [4]
      with the initial conditions  $T_0=0.57$ GeV, $\lambda_{g0}=0.09$ and
      $\lambda_{q0}=0.02$, $x_0=1.0$.
       (b)The time dependence of $T$, $x$, $\lambda_g$ and $\lambda_q$
       with initial conditions $T_0=0.57$ GeV, $\lambda_{g0}=0.09$ and
       $\lambda_{q0}=0.02$ for $x=0.0$, 1.0, 2.0. }
\label{Figure 3}
\end{figure}

 Fig. 3(a) shows temperature T and the fugacities as function of $\tau$ obtained
 with the revised values of $R_2$ and $R_3$ (solid curves). The dashed curves
 are obtained using Biro et al's factorization. Since $R_2$ values are found
 less as compared to Biro et al's result, the equilibration rate for $\lambda_q$
 slows down and the rate for $\lambda_g$ goes up marginally at large $\tau$.
 The equilibration rate for $x$ also slows down. This implies that the rate of decrease
(increase) of quark (anti-quark) contents becomes slower.
 However, as increase in the rate for $\lambda_g$ and decrease in the rate of $\lambda_q$ have
 opposite effects on temperature, the variation of T with $\tau$ remains practically
 unchanged. In Fig. 3(b), we
 show the effect of baryon density on chemical equilibration with
 revised rates $R_2$ and $R_3$. As before \cite{7}, the gluon equilibration rate
 slows down much more than the slight enhancement found in quark (anti-quark)
 equilibration rate. The overall effect is that the temperature of the plasma
 falls at a slower rate as compared to the plasma when it is baryon free.

\section{Thermal hard photon production}
   Hard  photons are the promising probes to study the evolution of the
   quark gluon plasma produced in relativistic heavy ion collisions.
   So far photon emission
   has been considered mostly from
   either chemically equilibrated or non-equilibrated plasma at zero
   baryon density \cite{14,15,16}. In Ref. \cite{17,18}, hard photon
   yield has been calculated from a chemically equilibrated
   baryon rich plasma.
   Here we investigate the photon production rate and integrated yields
  from a chemically unsaturated plasma at finite baryon density.

      In a thermodynamic system, the photon production rate and its momentum
  distribution depend on the momentum distribution of the quarks,
  anti-quarks and the gluons, which is governed by the thermodynamic state
  of the plasma. To get the emission rate of a photon from the plasma we have
  to fold the amplitude for these reactions with the thermal distribution
  functions and integrate over the phase space volume of all particles
  except photon .
 The calculation of photon emission rate from a QGP follows via two steps.
The emission rate from a stationary plasma at a fixed temperature T
is determined using Juttner distribution function
and  the yield is calculated by integrating over the plasma volume created
by the expansion and consequent cooling of the plasma.
  In thermal and chemical equilibrium, the production rate of hard
  photons with energy $E\gg T$ can be computed using the Braaten-Yaan
  prescription \cite{19}. Then the rate decomposes into a soft part, which is
  treated using a resummed quark propagator according to Braaten-Pisarsky
  Method \cite{20}, and a hard part containing only bare propagators and vertices.
  The medium effects in the QGP are included in the soft part through
  the thermal quark mass, which serves as an infrared cutoff in the case of
  a vanishing bare quark mass. The hard part follows from the momentum
  integration over the matrix elements that lead to photon emission in lowest
  order multiplied by the distribution functions of the incoming and
  outgoing partons \cite{14}. A separation scale $k_c$ is introduced, which allows one to
  distinguish between soft and hard momenta of the intermediate quark.

  In the following, we
 only consider the hard part of the photon emission rate and the cut-off
parameter $k_c^2$ is replaced by the thermal mass $2 m_q^2$.
  The main processes of photon production are ( O($\alpha \alpha_s$) annihilation
  $q {\bar q} \rightarrow g \gamma$ and Compton processes
  $q ({\bar q})g \rightarrow q ({\bar q}) \gamma$). The thermal rates for these
  reactions are
 \begin{eqnarray}
 2E \frac{dR^{hard}}{d^3p}&=&\frac {N}{(2 \pi)^8} \int \frac {d^3p_1}{2E_1}
    \frac {d^3p_2}{2E_2}  \frac {d^3p_3}{2E_3}~f_1(E_1)~f_2(E_2)~(1 \pm f_3(E_3))\nonumber \\
    && \delta (p_1^\mu +~  p_2^\mu -~  p_3^\mu -~ p^\mu )\sum |{\cal M}|^2
    \end{eqnarray}
 where $f_{1,2,3}$ are the parton  distribution function corresponding to two processes \cite{7},
 with plus sign for annihilation and the minus sign for the two Compton processes.
 Following Ref.\cite{14,18}, the above equation can be written as
 \begin{eqnarray}
 2E \frac{dR^{hard}}{d^3p}&=&\frac {1}{8~(2 \pi)^7~E} \int_{2k_c^2}^\infty~ds
 \int_{-s+k_c^2}^{-k_c^2}~dt~\sum |{\cal M}|^2\nonumber \\
 && \times \int_{R^2}~dE_1~dE_2 \frac{\Theta (P(E_1,E_2)) f_1 f_2 (1 \pm f_3)}
  {\sqrt {P(E_1,E_2)}}
    \end{eqnarray}
  where $P(E_1,E_2)=aE_2^2 + bE_2 + c$, with a,b,c as given by
  \equation
  a=-(s+t)^2~;~ b=2(s+t)(Es-E_2t)~;~ c=st(s+t)-(E_s+E_2)^2 \endequation
  and $\Theta$ is the step function, $s,~t,~u$ are the Mandelstam variables.
  The matrix elements for the Compton and
  annihilation processes can be written as
 \begin{eqnarray}
  \sum |{\cal M}|^2 &=& \frac {2^9~5}{9}~\pi^2~\alpha \alpha_s \frac{u^2+t^2}{ut}\nonumber \\
  \sum |{\cal M}|^2 &=&-\frac {2^9~5}{9}~\pi^2~\alpha \alpha_s \frac{s^2+t^2}{st}
    \end{eqnarray}

For the case $\mu=0$, it is possible to obtain an approximate analytic expression
for Eq.(45) assuming BM distribution for $f_1$ and $f_2$ and full Fermi/Bose
distribution functions for $f_3$ \cite{14}. Since we consider baryon rich
unsaturated plasma and  use Juttner distribution function,
we have to employ numerical methods. Following \cite{18}, we rewrite Eq.(45)
in a form suitable for
Gauss quadrature with $E_+ :=E_1+E_2$
 \begin{eqnarray}
 2E \frac{dR^{hard}}{d^3p}&=&\frac {5 \alpha \alpha_s}{18 \pi^5 E}
 e^{- E/T - k_c^2/2ET}
 \underbrace{\int_{2k_c^2}^\infty~~ds~~e^{-(s-2k_c^2)/4ET}}_{Lagauerre}~~ \frac{1}{s}~~
 \underbrace{\int_{-s+k_c^2}^{-k_c^2} dt }_{Legendre} |{\cal M}(s,t)|^2 ~~\nonumber\\
 && \underbrace{\int_{E+\frac{s}{4E}}^\infty
 dE_+~e^{-(E_+ -E-\frac{s}{4E})/T}}_{Lagauerre} ~.~\frac{1}{1 \mp \lambda_3~e^{-(E_+ -E)/T}} \nonumber\\
 &&\underbrace{\int_{E_2^-} ^{E_2^+} \frac{dE_2} {\sqrt{ P_1(E_+,E_2)}}}_{Chebyshev}
 ~~ \frac{\lambda_1} {1 \pm \lambda_1 e^{-(E_+-E_2)/T}}
 ~~ \frac{\lambda_2}{1 + \lambda_2 e^{-E_2/T}}
 \end{eqnarray}
 where the upper signs and
 $\lambda_1=\lambda_Q,~\lambda_2=\lambda_{\bar Q}~,\lambda_3=\lambda_g,~$
 is used for annihilation process, the Compton
 processes require lower signs and $\lambda_1=\lambda_g,
 ~\lambda_2=\lambda_3=\lambda_{Q({\bar Q})}$ and the two results for $\lambda_Q$ and
 $\lambda_{\bar Q}$ are to be added in order to take quarks and anti-quarks
 into consideration. The polynomial $P_1(E_+,E_2)$ in Eq.(48)
 is $P(E_+ -E_2,E_2)/s^2$ and has
 proper weight for Gauss-Chebyshev quadrature in $E_2$, where $E_2^-$
 and $E_2^+$ are the two roots of the polynomial $P(E_+,E_2)$. The $E_+$ integral
 as well as the s integral, is done numerically by Gauss-Laguerre quadrature
and the $t$ integral is performed using Gauss-Legendre quadrature. This way,
the accuracy of the integral has been improved. We have also verified the result
using BM distribution for $f_1$ and $f_2$ that matches with the analytic results
corresponding to equilibrated plasma at $x=0$ in the limit $k_c^2 \rightarrow 0$.
\begin{figure}[h!]
\vspace{3cm}
\hspace{3cm}
\begin{minipage}[t]{12cm}
\psfig{figure=fig4.out,width=12cm,height=12cm}
\end{minipage}
\caption{ (a) The photon production rate $2EdR/d^3p$
       for a chemically equilibrated plasma as a function of photon
       energy E at a fixed initial temperature $T_0= 0.265$ GeV
       corresponds to energy density $\epsilon=9.0$ GeV/fm$^3$ when $x=0$ for
       $x= 0.0$, 1.0, 1.5 and 2.0  and $\lambda_g=\lambda_q=1.0$.
       (b) The photon production rate $2EdR/d^3p$ as a function of photon
       energy E at a fixed initial energy density $\epsilon_0= 9.0$ GeV/fm$^3$ for
       $x= 0.0$, 1.0, 1.5 and 2.0  and $\lambda_g=\lambda_q=1.0$. }
\label{Figure 4}
\end{figure}

Figs. 4(a) and 4(b) show the photon production rate for a chemically equilibrated
plasma both at a fixed temperature ($T$=0.265 GeV) and fixed energy density
($\epsilon$=9.0 GeV/$fm^3$). The energy density $\epsilon$ of 9.0 Gev/fm$^3$ corresponds
to an initial temperature of 0.265 GeV at $x$=0. These results are similar
to what was found in Ref. \cite{18}. We have further extended the work to
calculate the photon production from
unsaturated plasma at finite baryon density. Figs. 5(a) and 5(b) shows the
corresponding results for an unsaturated plasma. The fugacities shown in the
figure caption correspond to an initial temperature of 0.57 GeV for energy
density $\epsilon$=9.0 GeV/$fm^3$ at $x=0$. Both Figs. 4(a) and 5(a) show
similar behaviour namely, that the rate of photon production is enhanced with
increasing baryon density. As will be shown below with analytic expression,
the annihilation part is not affected much by $x$ whereas the Compton
process for quark(anti-quark) is enhanced (suppressed) exponentially ($e^{\pm x}$).
Therefore, the enhancement with increasing baryon density comes due to
the quark part of the Compton process. At fixed energy density,
the temperature also decreases with the increasing baryon density. Hence the
net effect is that the photon rate decreases with increasing $x$ if the plasma is
saturated (see Fig. 4(b). In the case of unsaturated plasma, the decrease
of initial temperature with increasing $x$ is not very significant to compensate
the Compton enhancement. The overall effect is, therefore, that the rate still increases
with increasing $x$.

Since we are mainly interested to calculate the integrated photon yield
which involves a six-dimensional integration (see Eq. 52 below) and will be time consuming,
 we look for an approximate
analytic solution for photon production rate i.e. an approximation to Eq.(45).
For a chemically  unsaturated plasma, we can use MFD distribution functions
to factorize the product appearing in Eq. (44) as
  \equation
  f_1~f_2(1 \pm f_3)= \lambda_1\lambda_2\lambda_3 f_1^{eq}~f_2^{eq}~(1\pm f_3^{eq})
  +\lambda_1\lambda_2(1 -\lambda_3) f_1^{eq}~f_2^{eq}. \endequation
The factorization is similar to what has been used in Ref. \cite{16} to estimate
the photon yield from a baryon free  unsaturated plasma  except that the fugacities
$\lambda_1~,~\lambda_2~,~\lambda_3~$ are replaced by
$\lambda_g~,~\lambda_Q~,~\lambda_{\bar Q}$ corresponding to gluon, quark and
anti-quark respectively. The photon production rate can now be written in two
parts.
\begin{figure}[h!]
\hspace{3cm}
\begin{minipage}[t]{12cm}
\psfig{figure=fig5.out,width=12cm,height=12cm}
\end{minipage}
\caption{ (a) The photon production rate $2EdR/d^3p$
       chemically non-equilibrated plasma as a function of photon
       energy E at a fixed initial temperature $T_0= 0.265$ GeV for
       $x= 0.0$, 1.0, 1.5 and 2.0  and $\lambda_g=.09,~\lambda_q=.02$.
       The dots are the results of the approximate formula (sum of Eq.(51)
       Eq. (52)).
       (b) The photon production rate $2EdR/d^3p$ as a function of photon
       energy E at a fixed initial energy density $\epsilon_0= 9.0$ GeV/fm$^3$ for
       $x= 0.0$, 1.0, 1.5 and 2.0  and $\lambda_g=.09,~\lambda_q=1.0$.
       The dots are the results of the approximate formula (sum of Eq.(51)
       and Eq. (52)). }
\label{Figure 5}
\end{figure}

For the first part, one can use the analytic form that can be
obtained using BM distribution for $f_1$ and $f_2$ as in the case for equilibrated
baryon free plasma \cite{14} and given by
\begin{eqnarray}
\left(2E \frac{dR}{d^3p}\right)_1 &= &\frac{5\alpha~\alpha_s}{9~ \pi^2}~ T^2~e^{-E/T}
\Biggl [~\underbrace{~\lambda_Q \lambda_{\bar Q} \lambda_g \left \{
\frac{2}{3}~ln\left(\frac{4~E T}{k_c^2}\right)-1.43 \right\}}_{annihilation}\nonumber\\
 &&+~\underbrace{~(\lambda_Q^2 \lambda_g + \lambda_{\bar Q}^2~\lambda_g )\left\{
 \frac{1}{6} ln\left(\frac{4~E T}{k_c^2}\right)+0.0075 \right\}}_{Compton} \Biggr ]
\end{eqnarray}
Following Ref. \cite{16}, the second part can be written as
\begin{eqnarray}
\left(2E \frac{dR}{d^3p}\right)_2 &= &\frac{10\alpha~\alpha_s}{9~ \pi^4}~ T^2~e^{-E/T}
\Biggl [\underbrace{\lambda_Q \lambda_{\bar Q} (1 - \lambda_g)\Biggl \{- 2 -2 \gamma +
2~ln \left(\frac{4~ET}{k_c^2}\right)\Biggr\}}_{annihilation}\nonumber\\
&&+~\underbrace{\frac{1}{2}\left\{\lambda_Q \lambda_g (1- \lambda_Q) + \lambda_{\bar Q}
\lambda_g (1- \lambda_{\bar Q})\right \} \Biggl \{ 1 -2 \gamma +
2~ln \left(\frac{4~ET}{k_c^2}\right)\Biggr\}}_{Compton} \Biggr]
\end{eqnarray}
where $\alpha$, $\alpha_s$ are electromagnetic and strong coupling constants respectively.
In the above, the first term is the contribution from the annihilation whereas
the second and third terms are due to Compton like processes. The soft part
of the contribution is taken care by choosing the cut-off parameter $k_c^2=
2m_q^2$. As argued before, the annihilation part is nearly independent of $x$
(the dependence on $\mu$ through $k_c^2$ is logarithmic) where as
the Compton like processes depend on $x$ directly through $\lambda_Q$ or $\lambda_{\bar Q}$.
The photon rate is the sum of Eq.(50) and Eq.(51).
This formula reproduces the results of photon production rate quite
accurately (when the plasma is unsaturated) with the exact result
obtained from numerical integration. The results are shown in Figs. 5(a) and
5(b) (see solid circles) for comparison at two values of $x=0$ and $x=2.0$.
The above analytic expression also reproduces the result for saturated plasma
although the agreement is not so exact as in case of Figs. 5(a) and 5(b).

 Next we calculate  the photon yield by  integrating the above analytic expression
 of photon production rate (sum of Eq. (50) and Eq. (51))
over the space and time  \cite{16}
\equation
\left(\frac{2~dn}{d^2p_\perp~dy}\right)  = Q\int_{\tau_0}^{\tau_c}
d \tau \tau\int_{-y_{nuc}}^{y_{nuc}}dy^\prime\left(2E\frac{dn}{d^3p~d^4x}\right)
\left|_{E^{loc~rest}=p_\perp\cosh (y-y^\prime),T=T(\tau),\lambda=\lambda(\tau)}\right.
\endequation
The $\tau$ dependence of
 $T(\tau)$, $\lambda_g(\tau)$ and  $\lambda_{Q(\bar Q)}(\tau)$ are taken from
the above calculations (in previous section) with $\alpha_S$=0.3,
$\lambda_{g0}$ =0.09, $\lambda_{q0}$ =0.02 and $T_0=0.57$ GeV.
Further, we take
$y_{nuc}=\pm 6$ corresponding to RHIC energy. The transverse cross-section
of the gold nuclei, $Q\approx$ 180fm$^2$.
Figs. 6(a) and 6(b) show the  above integrated yields
both at a fixed initial temperature and fixed energy density. The integrated
yields both at fixed temperature and fixed energy density behave similarly
as the rates shown in  Figs. 5(a) and 5(b).
It may be pointed out that, the finite baryon density effects which result
in a slower cooling of the plasma affect  the thermal photon and dilepton production
yields somewhat differently. For example the basic rate of thermal dilepton production
(studied in previous work Ref. \cite{7}) does not depend on $x$ as it
involves only $\lambda_Q \lambda_{\bar Q}$.
\begin{figure}[h!]
\hspace{3cm}
\begin{minipage}[t]{12cm}
\psfig{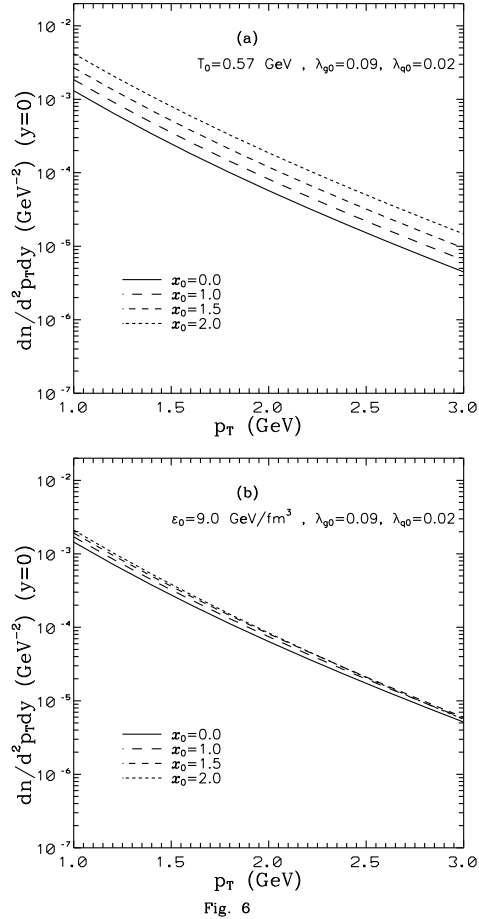}
\end{minipage}
\caption{ (a) The integrated photon yield  $dn/d^p_T~dy$ from a chemically non-equilibrated plasma
        at a fixed initial temperature $T_0=0.57$ GeV for various values of $x_0$.
        (b) The integrated photon yield from a chemically non-equilibrated plasma
        at a fixed initial energy density $\epsilon_0=9.0$ GeV/fm$^3$
        for various values of $x_0$. }
\label{Figure 6}
\end{figure}
But the integrated yield is enhanced
due to the slower space time evolution at finite baryon density. On the other hand,
in case of thermal photon production, the Compton like processes involve the
product of the type $\lambda_g \lambda_Q$ or $\lambda_g \lambda_{\bar Q}$. Therefore,
unlike the dilepton case, the basic photon yield also gets affected due to
chemical equilibration since the variation of $x$ and $\lambda_g$ are significant.

\section{ Summary and Conclusion }
 In the present work, we have investigated the effect of baryon density
 on parton production for the process $gg \rightleftharpoons ggg$ and
 $gg \rightleftharpoons q{\bar q}$ with full phase space distribution and
 also including quantum effects like Pauli blocking and Bose enhancement
 factors explicitly. The results   obtained using exact distribution function
 (Juttner) does not differ much from the results obtained using factorized
 distribution function of MFD or Boltzmann (BM) type. Although, BM approximation
 is classical, the final expression for the rate $gg \rightleftharpoons q{\bar q}$
 includes quantum statistics. The results which are obtained on the basis
 of Biro et al's estimation, although classical in nature, overestimate the
rate by 20$\%$ to 30$\%$ due to use of MFD type of approximation.
Further it is found that, the rates for both the processes
 ($gg \rightleftharpoons ggg$ and
 $gg \rightleftharpoons q{\bar q}$) are insensitive to baryon density particularly
 when the plasma is unsaturated. This further justifies the use of MFD or BM
type of distribution for partons for unsaturated plasma at small baryon density.
We have also studied the process of chemical equilibration using the revised
rates and also compare the results with the previous values obtained as per
Biro et al's estimation. It is found, that the quark and gluon equilibration
rates are affected slightly whereas the temperature remains unaffected.
We have also studied thermal hard photon yields from a chemically non-equilibrated
plasma which are considered to be the ideal probe to study the space time evolution of the
plasma. We employ numerical techniques to estimate photon rate using Juttner
distribution functions and also compare our results with approximate
analytic expression. It is seen that as in the case of the dilepton yields, the photon yield is also enhanced
in presence of finite baryon density.

\end{document}